\documentclass[number]{elsarticle}
\usepackage[british]{babel}
\usepackage{ucs}
\usepackage[utf8x]{inputenc}
\usepackage{amssymb,amsbsy,amsfonts,amsmath}
\usepackage{pifont}
\usepackage{pstricks,pst-node,pst-tree}
\usepackage{geometry}
\usepackage{graphicx}
\usepackage{hyperref}
\usepackage{color}

\newcommand{\noun}[1]{\textsc{#1}}

\providecommand{\boldsymbol}[1]{\mbox{\boldmath $#1$}}

\journal{Journal of Multivariate Analysis}

\DeclareMathOperator*{\indep}{{\;\bot\!\!\!\!\!\!\bot\;}} 
\DeclareMathOperator{\Dir}{Dir}

\newtheorem{theorem}{Theorem}

\newtheorem{corollary}[theorem]{Corollary}

\newtheorem{definition}[theorem]{Definition}
\newtheorem{example}[theorem]{Example}

\newtheorem{lemma}[theorem]{Lemma}

\newproof{proof}{Proof}

\begin{document}

\begin{frontmatter}

  \title{Bayesian MAP Model Selection of Chain Event Graphs}

  \author[warwick]{G.~Freeman} 
  \ead{g.freeman@warwick.ac.uk}

  \author[warwick]{J.Q.~Smith}
  \ead{j.q.smith@warwick.ac.uk}

  \address[warwick]{Department of Statistics, University of Warwick,
    Coventry, CV4 7AL}

\begin{abstract}
  The class of chain event graph models is a generalisation of the
  class of discrete Bayesian networks, retaining most of the
  structural advantages of the Bayesian network for model
  interrogation, propagation and learning, while more naturally
  encoding asymmetric state spaces and the order in which events
  happen. In this paper we demonstrate how with complete sampling,
  conjugate closed form model selection based on product Dirichlet
  priors is possible, and prove that suitable homogeneity assumptions
  characterise the product Dirichlet prior on this class of models. We
  demonstrate our techniques using two educational examples.
\end{abstract}
\end{frontmatter}

\section{Introduction} \label{sec:Introduction}

Bayesian networks (BNs) are currently one of the most widely used
graphical models for representing and analysing finite discrete
graphical multivariate distributions with their explicit coding of
conditional independence relationships between a system's variables
\cite{cowell_probabilistic_1999,lauritzen_graphical_1996}. However,
despite their power and usefulness, it has long been known that BNs
cannot fully or efficiently represent certain common scenarios. These
include situations where the state space of a variable is known to
depend on other variables, or where the conditional independence
between variables is itself dependent on the values of other
variables. Some examples of such latter scenarios are given by Poole
and Zhang \cite{poole_exploiting_2003}. In order to overcome such
deficiencies, enhancements have been proposed to the basic Bayesian
network in order to create so-called ``context-specific'' Bayesian
networks \cite{poole_exploiting_2003}. These have their own problems,
however: either they represent too much of the information about a
model in a non-graphical way, thus undermining the rationale for using
a graphical model in the first place, or they struggle to represent a
general class of models efficiently. Other graphical approaches that
seek to account for ``context-specific'' beliefs suffer from similar
problems. 

This has led to the proposal of a new graphical model ---
the chain event graph (\noun{CEGs}) --- which first propounded in
\cite{smith_conditional_2008}. As well as solving the aforementioned
problems associated with Bayesian networks and related graphical
models, CEGs are able, not unrelatedly, to encode far more efficiently
the common structure in which models are elicited --- as asymmetric
processes --- in a single graph. To this end, CEGs are based not on
Bayesian networks, but on event trees (ETs)
\cite{shafer_art_1996}. Event trees are trees where nodes represent
situations --- i.e.~scenarios in which a unit might find itself ---
and each node's extending edges represent possible future situations
that can develop from the current one. It follows that every atom of
the event space is encoded by exactly one root-to-leaf path, and each
root-to-leaf path corresponds to exactly one atomic event.  It has
been argued that ETs are expressive frameworks to directly and
accurately represent beliefs about a process, particularly when the
model is described most naturally, as in the example below, through
how situations might unfold \cite{shafer_art_1996}. However, as
explained in \cite{smith_conditional_2008}, ETs can contain excessive
redundancy in their structure, with subtrees describing
probabilistically isomorphic unfoldings of situations being
represented separately. They are also unable to explicitly express a
model's non-trivial conditional independences. The CEG deals with
these shortcomings by combining the subtrees that describe identical
subprocesses (see \cite{smith_conditional_2008} for further details),
so that the CEG derived from a particular ET has a simpler topology
while in turn expressing more conditional independence statements than
is possible through an ET.

We illustrate the construction and the types of symmetries it is
possible to code using a CEG with the following running example.

\begin{example} \label{ex:education-event-tree} Successful students on
  a one year programme study components $A$ and $B$, but not everyone
  will study the components in the same order: each student will be
  allocated to study either module $A$ or $B$ for the first 6 months
  and then the other component for the final 6 months. After the first
  6 months each student will be examined on their allocated module and
  be awarded a distinction (denoted with $D$), a pass ($P$) or a fail
  ($F$), with an automatic opportunity to resit the module in the last
  case. If they resit then they can pass and be allowed to proceed to
  the other component of their course, or fail again and be
  permanently withdrawn from the programme. Students who have
  succeeded in proceeding to the second module can again either fail,
  pass or be awarded a distinction. On this second round, however,
  there is no possibility of resitting if the component is
  failed. With an obvious extension of the labelling, this system can
  be depicted by the event tree given in Figure
  \ref{fig:education-event-tree}.

  \begin{figure}
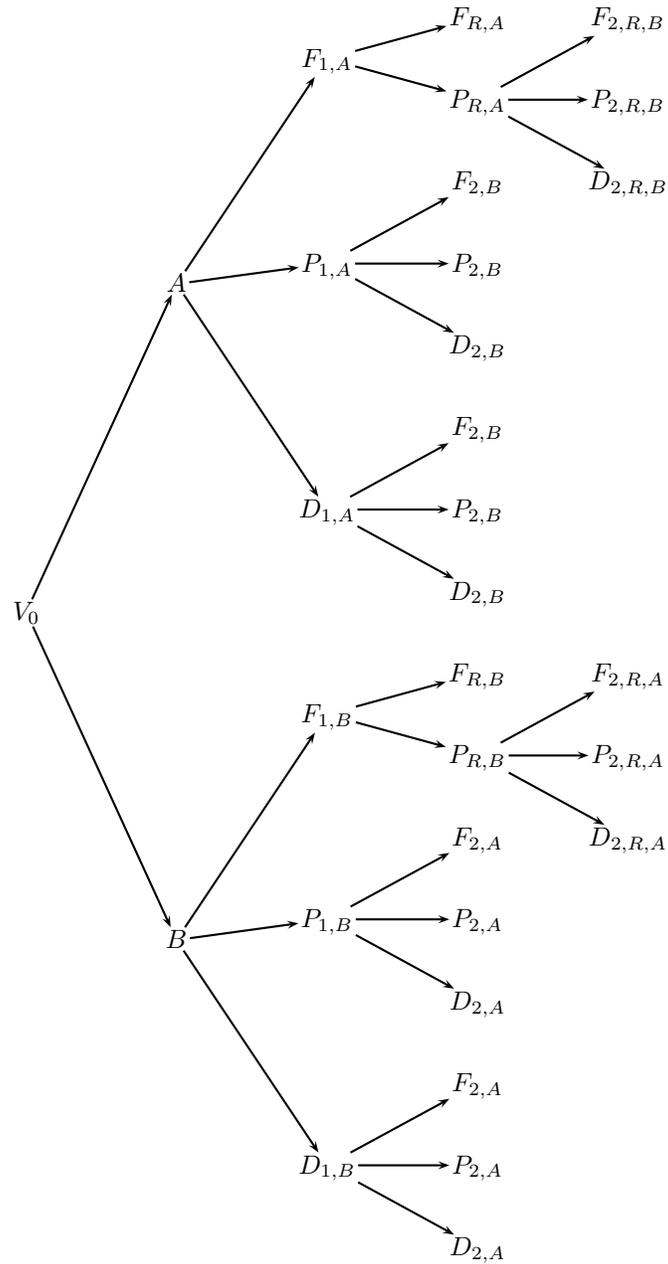

    \begin{center}

      \psset{arrows=->}
      \pstree[treemode=R,nodesep=1pt]{\TR{$V_0$}}{\pstree{\TR{$A$} } {
          \pstree{ \TR{$F_{1,A}$}}{{\TR{$F_{R,A}$}}
            \pstree{\TR{$P_{R,A}$}}{\TR{$F_{2,R,B}$}
              \TR{$P_{2,R,B}$}\TR{$D_{2,R,B}$}}}
          \pstree{\TR{$P_{1,A}$}}{\TR{$F_{2,B}$} \TR{$P_{2,B}$}
            \TR{$D_{2,B}$}} \pstree{\TR{$D_{1,A}$}}{\TR{$F_{2,B}$}
            \TR{$P_{2,B}$} \TR{$D_{2,B}$}} }
        \pstree{\TR{$B$}}{\pstree{\TR{$F_{1,B}$}}{ {\TR{$F_{R,B}$}}
            \pstree{\TR{$P_{R,B}$}}{\TR{$F_{2,R,A}$} \TR{$P_{2,R,A}$}
              \TR{$D_{2,R,A}$}} }
          \pstree{\TR{$P_{1,B}$}}{\TR{$F_{2,A}$} \TR{$P_{2,A}$}
            \TR{$D_{2,A}$}} \pstree{\TR{$D_{1,B}$}}{\TR{$F_{2,A}$}
            \TR{$P_{2,A}$} \TR{$D_{2,A}$}} } }

      \caption{Event tree of a student's potential progress through a
      hypothetical course described in Example
      \ref{ex:education-event-tree}. Each non-leaf node represents a
      juncture at which a random event will take place, with the
      selection of possible outcomes represented by the edges
      emanating from that node. Each edge distribution is defined
      conditional on the path passed through earlier in the tree to
      reach the specific node.}
  \end{center}
  \label{fig:education-event-tree}
  \end{figure}

\end{example}

To specify a full probability distribution for this model it is
sufficient to only specify the distributions associated with the
unfolding of each situation a student might reach. However, in many
applications it is often natural to hypothesise a model where the
distribution associated with the unfolding from one situation is
assumed identical to another. Situations that are thus hypothesised to
have the same transition probabilities to their children are said to
be in the same \emph{stage}. Thus in Example
\ref{ex:education-event-tree} suppose that as well as subscribing to
the ET of Figure \ref{fig:education-event-tree} we want to consider a
model also embodying the following three hypotheses:
\begin{enumerate}
\item The chances of doing well in the second component are the same
  whether the student passed first time or after a resit.
\item The components $A$ and $B$ are equally hard.
\item The distribution of marks for the second component is unaffected
  by whether students passed or got a distinction for the first
  component.
\end{enumerate}

These hypotheses can be identified with a partitioning of the
non-leaf nodes (situations). In Figure \ref{fig:education-event-tree} the set of
situations is 
\[
\mathcal{S}=\lbrace V_{0},A,B,P_{1,A},P_{1,B},D_{1,A},D_{1,B},F_{1,A},F_{1,B},P_{R,A},
P_{R,B}\rbrace .
\]
The partition $C$ of $\mathcal{S}$ that encodes exactly the above three
hypotheses consists of the stages $u_{1}=\left\lbrace  A,B\right\rbrace $,
$u_{2}=\left\lbrace  F_{1,A},F_{1,B}\right\rbrace  $, and
$u_{3}=\left\lbrace P_{1,A},P_{1,B},P_{R,A},P_{R,B},D_{1,A},D_{1,B
}\right\rbrace $
together with the singleton $u_{0}=\left\lbrace  V_{0}\right\rbrace $. Thus the
second stage $u_{2}$, for example, implies that the probabilities on
the edges $\left(F_{1,B},F_{R,B}\right)$ and
$\left(F_{1,A},F_{R,A}\right)$ are equal, as are the probabilities on
$\left(F_{1,B},P_{R,B}\right)$ and $\left(F_{1,A},P_{R,A}\right)$.
Clearly the joint probability distribution of the model -- whose atoms
are the root to leaf paths of the tree -- is determined by the
conditional probabilities associated with the stages. A CEG is the
graph that is constructed to encode a model that can be specified
through an event tree combined with a partitioning of its situations
into stages.

In this paper we suppose that we are in a context similar to that of
Example \ref{ex:education-event-tree}, where, for any possible model,
the sample space of the problem must be consistent with a single event
tree, but where on the basis of a sample of students' records we want
to select one of a number of different possible CEG models, i.e. we
want to find the ``best'' partitioning of the situations into
stages. We take a Bayesian approach to this problem and choose the
model with the highest posterior probability --- the Maximum A
Posteriori (MAP) model. This is the simplest and possibly most common
Bayesian model selection method, advocated by, for example, Dennison
et al \cite{denison_bayesian_2002}, Castelo
\cite{castelo_discrete_2002}, and Heckerman
\cite{heckerman_tutoriallearning_1999}, the latter two specifically
for Bayesian network selection. 

The paper is structured as follows. In the next section we review the
definitions of event trees and CEGs. In Section
\ref{sec:Conjugate-learning-of} we develop the theory of how conjugate
learning of CEGs is performed. In Section
\ref{sec:Local-Search-Algorithm-for-CEGs} we apply this theory by
using the posterior probability of a CEG as its score in a model
search algorithm that is derived using an analogous procedure to the
model selection of BNs. We characterise the product Dirichlet
distribution as a prior distribution for the CEGs' parameters under
particular homogeneity conditions. In Section \ref{sec:Actual-example}
the algorithm is used to discover a good explanatory model for real
students' exam results. We finish with a discussion.

\section{Definitions of event trees and chain event graphs}

In this section we briefly define the event tree and chain event
graph. We refer the interested reader to \cite{smith_conditional_2008}
for further discussion and more detail concerning their
construction. Bayesian networks, which will be referenced throughout
the paper, have been defined many times before. See
\cite{heckerman_tutoriallearning_1999} for an overview.

\subsection{Event Trees} \label{sec:ET-Definition}

Let $T=(V(T),E(T))$ be
a directed tree where $V(T)$ is its node set and $E(T)$ its edge
set. Let $S(T)=\lbrace v:v\in V(T)-L(T)\rbrace $ be the set of \noun{situations}
of $T$, where $L(T)$ is the set of \noun{leaf} (or \noun{terminal})
nodes. Furthermore, define $\mathbb{X}=\lbrace \lambda(v_{0},v):v\in
V(T)\backslash S(T)\rbrace $, where $\lambda(a,b)$ is the path from node $a$
to node $b$, and $v_{0}$ is the root node, so that $\mathbb{X}$ is the
set of root-to-leaf paths of $T$. Each element of $\mathbb{X}$ is
called an \noun{atomic event}, each one corresponding to a possible
unfolding of events through time by using the partial ordering induced
by the paths. Let $\mathbb{X}(v)$ denote the set of children of $v\in
V(T)$.  In an event tree, each situation $v\in S(T)$ has an associated
random variable $X(v)$ with sample space $\mathbb{X}(v)$, defined
conditional on having reached $v$. The distribution of $X(v)$ is
determined by the \noun{primitive probabilities}
$\lbrace \pi(v'|v)=p(X(v)=v'):v'\in\mathbb{X}(v)\rbrace $.  With random variables
on the same path being mutually independent, the joint probability of
events on a path can be calculated by multiplying the appropriate
primitive probabilities together. Each primitive probability
$\pi(v'|v)$ is a colour for the directed edge $e=(v,v')$, so that we
can have $\pi(e)=\pi(v'|v)$.

\begin{example}

  Figure \ref{fig:generic-simple-event-tree} shows a tree for two
  Bernoulli random variables, $X$ and $Y$, with $X$ occurring before
  $Y$. In an educational example $X$ could be the indicator variable
  of a student passing one module, and $Y$ the indicator variable for
  a subsequent module.

  \begin{figure}
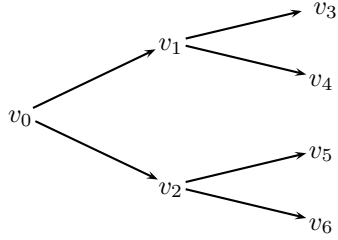


    \begin{center}
      \psset{arrows=->}
      \pstree[treemode=R,nodesep=1pt]{\TR{$v_0$}}{\pstree{\TR{$v_1$}}{\TR{
            $v_3$} \TR{$v_4$}} \pstree{\TR{$v_2$}}{\TR{$v_5$}
          \TR{$v_6$} } }

      \caption{Simple event tree. The non-zero-probability events in
        the joint probability distribution of two Bernoulli random
        variables, $X$ and $Y$, with $X$ observed before $Y$, can be
        represented by this tree. Here, all four joint states are
        possible, because there are four root-to-leaf paths through
        the nodes.} \label{fig:generic-simple-event-tree}

    \end{center}

  \end{figure}

  Here we have random variables $X(v_{0})=X$, $X(v_{1})=Y|(X=0)$ and
  $X(v_{2})=Y|(X=1)$, and primitive probabilities
  $\pi(v_{1}|v_{0})=p(X=0)$, $\pi(v_{3}|v_{1})=p(Y=0|X=0)$ and so on
  for every other edge. Joint probabilities can be found by
  multiplying primitive probabilities along a path,
  e.g.~$p(X=0,Y=0)=p(X=0)p(Y=0|X=0)=\pi(v_{1}|v_{0})\pi(v_{3}|v_{1})$
  as $v_{0}$ and $v_{1}$ are on a path.

\end{example}

\subsection{Chain Event Graphs}
\label{sec:CEG-Definition}

\begin{figure}
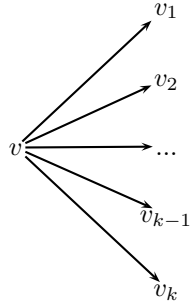


\begin{center}
  \psset{arrows=->}
  \pstree[treemode=R,nodesep=1pt]{\TR{$v$}}{\TR{$v_1$} \TR{$v_2$}
    \TR{...}  \TR{$v_{k-1}$} \TR{$v_k$}}

  \caption{Floret of $v$. This subtree represents both the random
    variable $X(v)$ and its state space
    $\mathbb{X}(v)$.} \label{fig:Floret-of-v}

\end{center}

\end{figure}

Starting with an event tree $T$, define a \noun{floret} of $v\in S(T)$
as \[
\mathcal{F}(v,T)=\left(V\left(\mathcal{F}(v,T)\right),E\left(\mathcal{F}(v,
    T)\right)\right)\] where $V(\mathcal{F}(v,T))=\lbrace v\rbrace \cup\lbrace v'\in
V(T):(v,v')\in E(T)\rbrace $ and $E(\mathcal{F}(v,T))=\lbrace e\in
E(T):e=(v,v')\rbrace $. The floret of a vertex $v$ is thus a sub-tree
consisting of $v$, its children, and the edges connecting $v$ and its
children, as shown in Figure \ref{fig:Floret-of-v}.  This represents,
as defined in section \ref{sec:ET-Definition}, the random variable
$X(v)$ and its sample space $\mathbb{X}(v)$.

One of the redundancies that can be eliminated from an ET is that of
the florets' edges of two situations, $v$ and $v'$ say, which have
identical associated edge probabilities despite being defined by
different conditioning paths. We say these two situations are at the
same \noun{stage}. This concept is formally defined as follows.

\begin{definition}
  Two situations $v,v'\in S(T)$ are in the same stage $u$ if and only
  if $X(v)$ and $X(v')$ have the same distribution under a bijection
  \[ \psi_{u}(v,v'):E(\mathcal{F}(v,T))\rightarrow
  E(\mathcal{F}(v',T))\] i.e. \[
  \psi_{u}(v,v'):\mathbb{X}(v)\rightarrow\mathbb{X}(v')\]

\end{definition} \label{def:stages}

The set of stages of an ET $T$ is written $J(T)$. This set partitions
the set of situations $S(T)$.

We can construct a \noun{staged tree}
$\mathcal{G}(T,L(T))$ with $V(\mathcal{G})=V(T)$,
$E(\mathcal{G})=E(T)$, and colour its edges
such that:

\begin{itemize}
\item If $v\in u$ and $u$ contains no other vertices, then all
  $(v,v^*)\in E(\mathcal{G})$ are left uncoloured;
\item If $v\in u$ and $u$ contains other vertices, then all
  $(v,v^*)\in E(\mathcal{G})$ are coloured; and
\item Whenever $e(v,v^*)\mapsto e(v',v'^*)$ under
  $\psi_{u}(v,v')$, then the two edges must have the
  same colour.
\end{itemize}
There is another type of situation that is of further interest. When
the whole development from two situations $v$ and $v'$ have identical
distributions, i.e.~there exists a bijection between their respective
subtrees similar to that between stages as defined in Definition
\ref{def:stages}, then the situations are said to be in the same
\noun{position}. This is defined formally as follows.

\begin{definition}
  Two situations $v,v'\in S(T)$ are in the same position $w$ if and
  only if there exists a bijection
  \[\phi_{w}(v,v'):\Lambda(v,T)\rightarrow\Lambda(v',T)\]
  where $\Lambda(v,T)$ is the set of paths in $T$ from $v$ to a leaf
  node of $T$, such that
  \begin{itemize}
  \item all edges in all of the paths in $\Lambda(v,T)$ and
    $\Lambda(v',T)$ are coloured in
    $\mathcal{G}(T,L(T))$; and
  \item for every path $\lambda(v)\in\Lambda(v,T)$, the ordered
    sequence of colours in $\lambda(v)$ equals the ordered sequence of
    colours in
    $\lambda(v'):=\phi_{w}(v,T)(\lambda(v))\in\Lambda(v',T)$
  \end{itemize}
\end{definition}

This ensures that when $v$ and $v'$ are in the same position, then
under the map $\phi_{w}(v,v')$ future development from either node
follows identical probability distributions.

We denote the set of positions as $K(T)$. Positions are an obvious way
of equating situations, because the different conditioning variables
of different nodes in the same position have no effect on any
subsequent development. It is clear that $K(T)$ is a finer partition
of $V(T)$ than $J(T)$, and indeed that $J(T)$ partitions $K(T)$, as
situations in the same position will also be in the same stage.

We now use stages and positions to compress the event tree into a chain
event graph. First, the \noun{probability graph} of the event tree
\[\mathcal{H}(\mathcal{G}(T))=\mathcal{H}(T)=(V(\mathcal{H}),E(\mathcal{H}))\]
is drawn, where $V(\mathcal{H})=K(T)\cup\lbrace w_{\infty}\rbrace $ and
$E(\mathcal{H})$ is constructed as follows.
\begin{itemize}
\item For each pair of positions $w,w'\in K(T)$, if there exists
  $v,v'\in S(T)$ such that $v\in w$,$v'\in w'$ and $e(v,v')\in E(T)$,
  then an associated edge $e(w,w')\in E(\mathcal{H})$ is
  drawn. Furthermore, if for a position $w$ there exists $v\in S(T)$,
  $v'\in L(T)$ and $e(v,v')\in E(T)$ such that $v\in w$, then an associated edge $e(w,w_\infty)\in E(\mathcal{H})$
  is drawn.
\item The colour of this edge, $e(w,w')$, is the same as the colour of
  the associated edge $e(v,v')$.
\end{itemize}

Now the CEG can finally be constructed by taking the probability graph
$\mathcal{H}(T)$ and connecting the positions that are in the same
stage using undirected edges: Let $\mathcal{C}(T)$ be a mixed graph
with vertex set $V(\mathcal{C})=V(\mathcal{H})$, directed edge set
$E_{d}(\mathcal{C})=E(\mathcal{H})$, and undirected edge set
$E_{u}(\mathcal{C})=\lbrace (w,w'):u(w)=u(w'),\, w,w'\in V(\mathcal{C})\rbrace $.

An example of a CEG that could be constructed from the event tree in
Figure \ref{fig:education-event-tree} is shown in Figure
\ref{fig:education-ceg-with-data}.

\section{Conjugate learning of CEGs} \label{sec:Conjugate-learning-of}

One convenient property of CEGs is that conjugate updating of the
model parameters proceeds in a closely analogous fashion to that on a
BN. Conjugacy is a crucial part of the model selection algorithm that
will be described in Section
\ref{sec:Local-Search-Algorithm-for-CEGs}, because it leads to closed
form expressions for the posterior probabilities of candidate
CEGs. This in turn makes it possible to search the often very large
model space quickly to find optimal models. We demonstrate here how a
conjugate analysis on a CEG proceeds.

Let a CEG $C$ have set of stages $J(C)=\lbrace u_{1},\dots,u_{k}\rbrace $,
and let each stage $u_{i}$ have $k_{i}$ emanating edges (labelled
$e_{1},\dots,e_{k_{i}}$) with associated probability vector
$\boldsymbol{\pi}_{i}=(\pi_{i1},\pi_{i2},\ldots,\pi_{ik_{i}})^{\prime}$
(where $\sum_{j=1}^{k_{i}}\pi_{ij}=1$ and $\pi_{ij}>0$ for
$j\in\lbrace1,\dots,k\rbrace$).  Then, under random sampling, the likelihood of
the CEG can be decomposed into a product of the likelihood of each
probability vector, i.e.
\[
p(\boldsymbol{x}|\boldsymbol{\pi},C)=\prod\limits
_{i=1}^{k}p_i(\boldsymbol{x}_{i}|\boldsymbol{\pi}_{i},C)
\] 
where $\boldsymbol{\pi}=\left\lbrace
  \boldsymbol{\pi}_{1},\boldsymbol{\pi}_{2},\ldots,
  \boldsymbol{\pi}_{k}\right\rbrace $, and
$\boldsymbol{x}=\left\lbrace
  \boldsymbol{x}_{1},\dots,\boldsymbol{x}_{k}\right\rbrace$ is the
complete sample data such that each
$\boldsymbol{x}_{i}=(x_{i1},\dots,x_{ik_{i}})^{\prime}$ is the vector
of the number of units in the sample (for example, the students in
Example 1) that start in stage $u_{i}$ and move to the stage at the
end of edge $e_{ij}$ for $j\in\lbrace 1,\dots,k_{i}\rbrace$.

If it is further assumed that $\boldsymbol{x}_i\indep\boldsymbol{x}_j
| \boldsymbol{\pi} ,\forall i\neq j $ then
\begin{equation}
p_i(\boldsymbol{x}_{i}|\boldsymbol{\pi}_{i},C)=\prod\limits
_{j=1}^{k_{i}}\pi_{ij}^{x_{ij}} \label{eq:multinomial-likelihood}
\end{equation}
Thus, just as for the analogous situation with BNs, the likelihood of
a random sample also separates over the components of
$\boldsymbol{\pi}$. With BNs, a common modelling assumption is of
local and global independence of the probability parameters
\cite{spiegelhalter_sequential_1990}; the corresponding assumption
here is that the parameters
$\boldsymbol{\pi}_{1}$,$\boldsymbol{\pi}_{2}$,$\ldots$,$\boldsymbol{\pi}_{k}$
of $\boldsymbol{\pi}$ are all mutually independent a priori. It will
then follow, with the separable likelihood, that they will also be
independent a posteriori.

If the probabilities $\boldsymbol{\pi}_{i}$ are assigned a Dirichlet
distribution, $\Dir (\boldsymbol{\alpha}_{i})$, a priori, where
$\boldsymbol{\alpha}_{i}=(\alpha_{i1},\alpha_{i2},\ldots,\alpha_{ik_{i}})^{\prime}$,
so that for values of $\pi_{ij}$ such that
$\sum_{j=1}^{k_{i}}\pi_{ij}=1$ and $\pi_{ij}>0$ for $1\leq j\leq
k_{i}$, the density of $\boldsymbol{\pi}_i$, $q_i(\boldsymbol{\pi}_i|C)$, can be written
\[
q_i(\boldsymbol{\pi}_{i}|C)=\frac{\Gamma(\alpha_{i1}+\ldots+\alpha_{ik_{i}})}{
  \Gamma(\alpha_{i1})\ldots\Gamma(\alpha_{ik_{i}})}\prod\limits
_{j=1}^{k_{i}}\pi_{ij}^{\alpha_{ij}-1}
\]
where $\Gamma(z)=\int_0^\infty t^{z-1} e^{-t} dt$ is the Gamma
function. It then follows that
$\boldsymbol{\pi}_{i}|\boldsymbol{x}$
$(=\boldsymbol{\pi}_{i}|\boldsymbol{x}_i)$ also has a Dirichlet
distribution, $\Dir (\boldsymbol{\alpha}_{i}^{\ast})$, a posteriori, where
$\boldsymbol{\alpha}^*_i=(\alpha_{i1}^*,\dots,\alpha_{ik_{i}}^*)^{\prime}$,
$\alpha_{ij}^{\ast}=\alpha_{ij}+x_{ij}$ for $1\leq j\leq k_{i},1\leq
i\leq k$.  The marginal likelihood of this model can be written down
explicitly as the function of the prior and posterior Dirichlet
parameters:
\[ 
p(\boldsymbol{x}|C) = \prod_{i=1}^{k}\left[\frac{\Gamma(\sum_{j}\alpha_{ij})}{\Gamma(\sum_{j}\alpha_{
        ij}^{*})}\prod_{j=1}^{k_{i}}\frac{\Gamma(\alpha_{ij}^{*})}{\Gamma(\alpha_{ij})}\right]
  \label{eq:Marginal-likelihood}.
\]
The computationally more useful logarithm of the marginal likelihood is therefore a linear
combination of functions of $\alpha_{ij}$ and $\alpha_{ij}^{*}$. Explicitly,
\begin{equation} 
  \log p(\boldsymbol{x}|C) = \sum_{i=1}^{k}{\left[s(\boldsymbol{\alpha}_{i})-s(\boldsymbol{\alpha}_{i}^{\ast})\right]} + \sum_{i=1}^{k}{\left[t(\boldsymbol{\alpha}_{i}^{\ast})-t(\boldsymbol{\alpha}_{i})\right]} \label{eq:marginal-likelihood}
\end{equation}
where for any vector $\mathbf{c}=(c_1,c_2,\dots,c_n)^\prime$,
\begin{equation}
  s(\mathbf{c})=\log\Gamma(\sum_{v=1}^n{c_{v}})
  \mbox{ and }
  t(\mathbf{c})=\sum_{v=1}^n{\log\Gamma(c_{v})}   \label{eq:s-and-t}
\end{equation}

So the posterior probability of a CEG $C$ after observing $\boldsymbol{x}$, $q(C|\boldsymbol{x})$, can
be calculated using Bayes' Theorem, given a prior probability $q(C)$:
\begin{equation}
\log q(C|\boldsymbol{x}) = \log p(\boldsymbol{x}|C) + \log q(C) + K \label{eq:posteriorC}
\end{equation}
for some value $K$ which does not depend on $C$. This is the
\noun{score} that will be used when searching over the candidate set
of CEGs for the model that best describes the data.

\section{A Local Search Algorithm for Chain Event Graphs}
\label{sec:Local-Search-Algorithm-for-CEGs}

\subsection{Preliminaries}
\label{sec:preliminaries}

With the log marginal posterior probability of a CEG model, $\log q(C|\boldsymbol{x})$, as its score,
searching for the highest-scoring CEG in the set of all candidate
models is equivalent to trying to find the Maximum A Posteriori (MAP)
model \cite{bernardo_bayesian_1994}. The intuitive approach for
searching $\boldsymbol{C}$, the candidate set of CEGs --- calculating
$q(C|\boldsymbol{x})$ (or $\log q(C|\boldsymbol{x})$) for every $C \in
\boldsymbol{C}$ and choosing $C^*:=\max_C q(C|\boldsymbol{x})=\max_C
\log q(C|\boldsymbol{x})$ --- is infeasible for any but the most
trivial problems. We describe in this section an algorithm for
efficiently searching the model space by reformulating the model
search problem as a clustering problem.

As mentioned in Section \ref{sec:CEG-Definition}, every CEG that can
be formed from a given event tree can be identified exactly with a
partition of the event tree's nodes into stages.  The coarsest
partition $C_{\infty}$ has all nodes with $k$ outgoing edges in the
same stage, $u_{k}$; the finest partition $C_0$ has each situation in
its own stage, except for the trivial cases of those nodes with only
one outgoing edge. Defined this way, the search for the
highest-scoring CEG is equivalent to searching for the highest-scoring
clustering of stages.

Various Bayesian clustering algorithm exist \cite{lau_bayesian_2007},
including many involving MCMC \cite{richardson_bayesian_1997}. We show
here how to implement an Bayesian agglomerative hierarchical
clustering (AHC) exact algorithm related to that of Heard et al
\cite{heard_quantitative_2006}. The AHC algorithm here is a local
search algorithm that begins with the finest partition of the nodes of
the underlying ET model (called $C_0$ above and henceforth) and seeks
at each step to find the two nodes that will yield the highest-scoring
CEG if combined.

Some optional steps can be taken to simplify the search, which we will
implement here. The first of these involves the calculation of the
scores of the proposed models in the algorithm. By assuming that the
probability distributions of stages that are formed from the same
nodes of the underlying ET are equal in all CEGs,
i.e. $p(\boldsymbol{x_{i}}|\boldsymbol{\pi_{i}},C_1)=p(\boldsymbol{x_{i}}|\boldsymbol{\pi_{i}},C_2),
\forall C_1,C_2\in \boldsymbol{C}$, it becomes more efficient to
calculate the differences of model scores, i.e.~the logarithms of the
relevant Bayes factors, than to calculate the two individual model
scores absolutely. This is because, if for two CEGs their stage sets
$J(C_1)$ and $J(C_2)$ differ only in that stages $u_{1a},u_{1b}\in
C_1$ are combined into $u_{2c}\in C_2$, with all other stages
unchanged, then the calculation of the logarithm of their posterior
Bayes factor depends only on the stages involved; using the notation
of Equation (\ref{eq:s-and-t}),
\begin{align}
  \log{\frac{q(C_1|\boldsymbol{x})}{q(C_2|\boldsymbol{x})}} & = \log{q(C_1|\boldsymbol{x})}-\log{q(C_2|\boldsymbol{x})} \\
  & = \log{q(C_1)}-\log{q(C_2)}+\log{q(\boldsymbol{x}|C_1)}-\log{q(\boldsymbol{x}|C_2)}\\
  \begin{split}    
    & =\log{q(C_1)}-\log{q(C_2)}+\sum_{i}{\left[s(\boldsymbol{\alpha}_{1i})-s(\boldsymbol{\alpha}_{1i}^{\ast})\right]} + \sum_{i}{\left[t(\boldsymbol{\alpha}_{1i}^{\ast})-t(\boldsymbol{\alpha}_{1i})\right]} \\
    & \qquad {}- \sum_{i}{\left[s(\boldsymbol{\alpha}_{2i})-s(\boldsymbol{\alpha}_{2i}^{\ast})\right]} - \sum_{i}{\left[t(\boldsymbol{\alpha}_{2i}^{\ast})-t(\boldsymbol{\alpha}_{2i})\right]} 
  \end{split} \\
  \begin{split}
    & = \log{q(C_1)}-\log{q(C_2)}+s(\boldsymbol{\alpha}_{1a})-s(\boldsymbol{\alpha}_{1a}^*)+t(\boldsymbol{\alpha}_{1a}^*)-t(\boldsymbol{\alpha}_{1a}) \\
    & \qquad {}+ s(\boldsymbol{\alpha}_{1b})-s(\boldsymbol{\alpha}^*_{1b})+t(\boldsymbol{\alpha}^*_{1b})-t(\boldsymbol{\alpha}_{1b}) \\
    & \qquad \qquad {}-s(\boldsymbol{\alpha}_{2c})+s(\boldsymbol{\alpha}_{2c}^*)-t(\boldsymbol{\alpha}_{2c}^*)+t(\boldsymbol{\alpha}_{2c})
  \end{split}
  \label{eq:log-Bayes-factor}
\end{align}
Using the trivial result that for any three CEGs
\begin{equation*}
\log q(C_3|\boldsymbol{x}) - \log q(C_2|\boldsymbol{x}) = \left[ \log
q(C_3|\boldsymbol{x}) - \log q(C_1|\boldsymbol{x}) \right] - \left[ \log
q(C_2|\boldsymbol{x}) - \log q(C_1|\boldsymbol{x}) \right],
\end{equation*}
it can be seen that in the course of the AHC algorithm, comparing two
proposal CEGs from the current CEG can be done equivalently by
comparing their log Bayes factors with the current CEG, which as shown
above requires fewer calculations.

The calculation of the score for each CEG $C$, as shown by Equation
(\ref{eq:posteriorC}), shows that it is formed of two components: the
prior probability of the CEG being the true model and the marginal
likelihood of the data. These must therefore be set before the
algorithm can be run, and it is here that the other simplifications
are made.

\subsection{The prior over the CEG space}
\label{sec:ceg-prior}

For any practical problem $\boldsymbol{C}$, the set of all possible
CEGs for a given ET, is likely to be a very large set, making setting
a value for $q(C),\forall C\in \boldsymbol{C}$ a non-trivial task. An
obvious way to set a non-informative or exploratory prior is to choose
the uniform prior, so that $q(C)=\frac{1}{\left| \boldsymbol{C}
  \right| }$. This has the advantages of being simple to set and of
eliminating the $\log{q(C_1)}-\log{q(C_2)}$ term in Equation
(\ref{eq:log-Bayes-factor}).

A more sophisticated approach is to consider which potential clusters
are more or less likely a priori, according to structural or causal
beliefs, and to exploit the modular nature of CEGs by stating that the
prior log Bayes factor of a CEG relative to $C_0$ is the sum of the
prior log Bayes factors of the individual clusters relative to their
components completely unclustered, and that these priors are modular
across CEGs. This approach makes it simple to elicit priors over
$\boldsymbol{C}$ from a lay expert, by requiring the elicitation only
of the prior probability of each possible stage.

A particular computational benefit of this approach is when the prior
Bayes factor of any CEG $C$ with $C_0$ is believed to be
zero, because one or more of its clusters is considered to be
impossible. This is equivalent in the algorithm to not including the
CEG in its search at all, as though it was never in $\boldsymbol{C}$ in
the first place, with the obvious simplification of the search following.

\subsection{The prior over the parameter space}
\label{sec:parameter-prior}

Just as when attempting to set $q(C)$, the size of most CEGs in
practise leads to intractability of setting $p(\boldsymbol{x}|C)$ for
each CEG $C$ individually. However, the task is again made possible by
exploiting the structure of a CEG with judicious modelling
assumptions.

Assuming independence between the likelihoods of the stages for every
CEG, so that $p(\boldsymbol{x}|\boldsymbol{\pi},C)$ is as determined
by Equation (\ref{eq:multinomial-likelihood}), and the fact that
$p(\boldsymbol{x}|C)=\int
p(\boldsymbol{x}|\boldsymbol{\pi},C)p(\boldsymbol{\pi}|C)d\boldsymbol{\pi}$,
it is clear that to set the marginal likelihood for each CEG is
equivalent to setting the prior over the CEG's parameters,
i.e.~setting $p(\boldsymbol{\pi}|C)$ for each $C$. With the two
further structural assumptions that the stage priors are independent
for all CEGs (so that $p(\boldsymbol{\pi}|C)=\prod_{i=1}^k
p(\boldsymbol{\pi}_i|C)$) and that equivalent stages in different CEGs
have the same prior distributions on their probability vectors, 
(i.e.~$p(\boldsymbol{\pi}_i|C_1)=p(\boldsymbol{\pi}_i|C_2)$), it can
be seen that the problem of setting
$p(\boldsymbol{x}|\boldsymbol{\pi},C)$ is reduced to setting the
parameter priors of each non-trivial floret in $C_0$
($p(\boldsymbol{\pi}_i|C_0),i=1,\dots,k$) and the parameter priors of
stages that are clusters of stages of $C_0$.

The usual prior put on the probability parameters of finite discrete
BNs is the product Dirichlet distribution. In Geiger and Heckerman
\cite{geiger_characterization_1997} the surprising result was shown
that a product Dirichlet prior is inevitable if local and global
independence are assumed to hold over all Markov equivalent graphs on
at least two variables. In this paper we show that a similar
characterisation can be made for CEGs given the assumptions in the
previous paragraph. We will first show that the floret parameters in
$C_0$ must have Dirichlet priors, and second that all CEGs formed by
clustering the florets in $C_0$ have Dirichlet priors on the stage
parameters. One characterisation of $C_0$ is given by Theorem
\ref{theorem:dirichlet-for-c0}.

\begin{theorem}
\label{theorem:dirichlet-for-c0}
  If it is assumed a priori that the rates at which units take the
  root-to-leaf paths in $C_0$ are independent (``path independence'')
  and that the probability of which edge units take after arriving at
  a situation $v$ is independent of the rate at which units arrive at
  $v$ (``floret independence''), then the non-trivial florets of $C_0$ have
  independent Dirichlet priors on their probability vectors.
\end{theorem}

\begin{proof}
  The proof is in the Appendix.
\end{proof}

Thus $p(\boldsymbol{\pi}_i|C_0)$ is entirely determined by the stated
rates $\gamma(\lambda)$ on the root-to-leaf paths
$\lambda\in\Lambda(C_0)$ of $C_0$. This is similar to the ``equivalent
sample sizes'' method of assessing prior uncertainty of Dirichlet
hyperparameters in BNs as discussed in Section 2 of Heckerman
\cite{heckerman_tutoriallearning_1999}.

Another way to show that all non-trivial situations in $C_0$ have
Dirichlet priors on their parameter spaces is to use the
characterisation of the Dirichlet distribution first proven by
Geiger and Heckerman \cite{geiger_characterization_1997}, repeated
here as Theorem \ref{theorem:geiger-and-heckerman}.

\begin{theorem}
  \label{theorem:geiger-and-heckerman}
  Let $\lbrace \theta_{ij}\rbrace ,1\leq i\leq k, 1\leq j\leq n,
  \sum_{ij}{\theta_{ij}} = 1$, where $k$ and $n$ are integers greater
  than 1, be positive random variables having a strictly positive pdf
  $f_U(\lbrace \theta_{ij}\rbrace )$. Define
  $\theta_{i.}=\sum_{j=1}^{n}{\theta_{ij}}$,
  $\theta_{I.}=\lbrace \theta_{i.}\rbrace _{i=1}^{k-1}$,
  $\theta_{j|i}=\theta_{ij}/{\sum_j{\theta_{ij}}}$, and
  $\theta_{J|i}=\lbrace \theta_{j|i}\rbrace _{j=1}^{n-1}$.

  Then if $\lbrace \theta_{I.},\theta_{J|1},\dots,\theta_{J|k}\rbrace $ are
  mutually independent, $f_U(\lbrace \theta_{ij}\rbrace )$ is Dirichlet.
\end{theorem}
\begin{proof}
  Theorem 2 of Geiger and Heckerman \cite{geiger_characterization_1997}.
\end{proof}

\begin{corollary}
  \label{corrolary:dirichlet-tree-from-geiger-and-heckerman}
  If $C_0$ has a composite number $m$ of root-to-leaf paths and all
  Markov equivalent CEGs have independent floret distributions then
  the vector of probabilities on the root-to-leaf paths of $C_0$ must
  have a Dirichlet prior. This means in particular that, from the
  properties of the Dirichlet distribution, the floret of each
  situation with at least two outgoing edges has a Dirichlet prior on
  its edges.
\end{corollary}

\begin{proof}
  Construct an event tree $C_0^\prime$ with $m$ root-to-leaf paths,
  where the floret of the root node $v_0^\prime$ has $k$ edges and
  each of the florets extending from the children of $v_0^\prime$ have
  $n$ edges terminating in leaf nodes, where $m=kn,k\geq 2,n\geq
  2$. This will always be possible with a composite $m$. $C_0^\prime$
  describes the same atomic events as $C_0$ with a different decomposition.

  Let the random variable associated with the root floret of $C_0^\prime$ be $X$, and
  let the random variable associated with each of the other florets be
  $Y|X=i,i=1,\dots,k$. Let $\theta_{ij}=P(X=i,Y=j)$. Then by the
  definition of event trees, $P(\theta_{ij}>0)>0,1\leq i\leq k,1\leq j\leq
  n$ and $\sum \theta_{ij} = 1$. By the notation of Theorem
  \ref{theorem:geiger-and-heckerman}, $\theta_{i.}=P(X=i)$ and
  $\theta_{j|i}=P(Y=j|X=i)$.

  By hypothesis the floret distributions of $C_0^{\prime}$ are
  independent. Therefore the condition of Theorem
  \ref{theorem:geiger-and-heckerman} holds and hence
  $f_U(\theta_{ij})$ is Dirichlet. From the equivalence of the atomic
  events, the probability distribution over the root-to-leaf path
  probabilities of $C_0$ is also Dirichlet, and so by Lemma
  \ref{lemma:floret-is-dirichlet}, all non-trivial florets of $C_0$
  therefore have Dirichlet priors on their probability vectors.
\end{proof}

To show that the stage parameters of all the other CEGs in
$\boldsymbol{C}$ have independent Dirichlet priors, an inductive
approach will be taken. Because of the assumption of consistency --
that two identically composed stages in different CEGs have identical
priors on their parameter space -- for any given CEG $C$ whose stages
all have independent Dirichlet priors on their parameters spaces, it
is known that another CEG $C^*$ formed by clustering two stages
$u_{1c},u_{2c}$ from $C$ into one stage $u_{c^{*}}$ will have
independent Dirichlet priors on all its stages apart from
$u_{c^{*}}$. It is thus only required to show that
$\boldsymbol{\pi}_{c^{*}}$ has a Dirichlet prior. We prove this result
for a class of CEGs called \noun{regular CEGs}.

\begin{definition}
  A stage $u$ is \noun{regular} if and only if every path
  $\lambda\in\Lambda(C)$ contains either one situation in $u$ or none
  of the situations in $u$.
\end{definition}

\begin{definition}
  A CEG is \noun{regular} if and only if every situation $u \in
  \boldsymbol{u}(C)$ is regular.
\end{definition}

\begin{theorem}
  \label{theorem:dirichlet-for-combination-stage}
  Let $C$ be a regular CEG, and let $C^*$ be the CEG that is formed
  from $C$ by setting two of its stages, $u_{1c}$ and $u_{2c}$, as
  being in the same stage $u_{c^{*}}$, where $u_{c^{*}}$ is a regular
  stage, with all other attributes of the CEG unchanged from $C$.

  If all stages in $C$ have Dirichlet priors, then assuming that
  equivalent stages in different CEGs have equivalent priors, all
  stages in $C^{*}$ have Dirichlet priors.
\end{theorem}

\begin{proof}
  Without loss of generality, let all situations in $u_{1c}$ and
  $u_{2c}$ have $s$ children each, and let the total number of
  situations in $u_{1c}$ and $u_{2c}$ be $r$. Thus there are $r$
  situations in $u_{c^{*}}$, each with $s$ children. By the assumption
  of prior consistency across stages, all stages in $C^{*}$ have
  Dirichlet priors on their parameter spaces, so it is only required
  to prove that $u_{c^{*}}$ has a Dirichlet prior.

  Consider the CEG $C^\prime$ formed as follows: Let the root node of
  $C^\prime$, $v_0$, have 2 children, $v_1$ and $v^\prime$. Let
  $v^\prime$ be a terminal node, and let $v_1$ have $r$ children,
  $\lbrace v_1(1),\dots,v_1(r)\rbrace $, which are equivalent to the situations in
  $u_{c^{*}}$, including the property that they are in the same
  stage $u_{c^\prime}$. Lastly, let the children of $\lbrace v_1(1),\dots,v_1(r)\rbrace $,
  $\lbrace v_1(1,1),\dots,v_1(1,s),\dots,v_1(r,1),\dots,v_1(r,s)\rbrace $, be leaf
  nodes in $C^\prime$.

  
  By construction, the prior for $u_{c^\prime}$ is the same as that
  for $u_{c^*}$.

  Now construct another CEG $C^{*\prime}$ from $C^\prime$ by reversing
  the order of the stages $v_1$ and $u_{c^\prime}$. The new CEG has
  root node $v_0$ with the same distribution as $v_0\in
  C^\prime$. $v_0$ now has two children $v^\prime$ -- the same as
  before -- and $v_2$, which has $s$ children
  $\lbrace v_2(1),\dots,v_2(s)\rbrace $ in the same stage. Each node $v_2(i),i=1,\dots,s$ has $r$
  children $v_2(i,1),\dots,v_2(i,r)$, all of which are leaf nodes.

  The two CEGs $C^{*\prime}$ and $C^\prime$ are Markov equivalent, as
  it is clear that
  $P(v_1(i,j))=P(v_2(j,i)),i=1,\dots,r,j=1,\dots,s$. The probabilities
  on the floret of $v_2$ are thus equal to the probabilities of the
  situations in the stage of $u_{c^\prime}$, and hence
  $u_{c^{*}}$. Because $v_2$ is a stage with only one situation,
  Theorem \ref{theorem:dirichlet-for-c0} implies that it has a
  Dirichlet prior. Therefore $u_{c^{*}}$ has a Dirichlet prior.
\end{proof}

An alternative justification for assigning a Dirichlet prior to any
stage that is formed by clustering situations with Dirichlet priors on
their state spaces can be obtained which does not depend on assuming
Markov equivalency between CEGs derived from different event trees by
assuming a property analogous to that of ``parameter modularity'' for
BNs \cite{heckerman_bayesian_1995}. This property states that the
distribution over structures common to two CEGs should be identical.

\begin{definition}
\label{def:margin-equivalency-for-stages}
  Let $u$ be a stage in a CEG $C$ composed of the situations
  $v_1,\dots,v_n$ from $C_0$, each of which has $m$ children
  $v_{i1},\dots,v_{im},i=1,\dots,n$ such that $v_{ij}$ are the same
  colour for all $i$ for each $j$. Then $u$ has the property of
  \noun{margin equivalency} if
  \begin{align}
  \pi_{uj}&=P(v_{1j} \mbox{ or } v_{2j} \mbox{ or } \dots \mbox{ or }
  v_{nj} | v_1 \mbox{ or } v_2 \mbox{ or } \dots \mbox{ or } v_n) \\
  &=\frac{\sum_{i=1}^n{P(v_{ij})}}{\sum_{i=1}^n{P(v_i)}}
  \end{align}
  is the same for both $C$ and $C_0$ for $j=1,\dots,m$.
\end{definition}

\begin{definition}
  $C$ has margin equivalency if all of its stages have margin equivalency.
\end{definition}

\begin{theorem}  
\label{theorem:stages-are-dirichlet-by-margin-equivalency}
  Let $u_{c}$ be a stage as defined in Definition
  \ref{def:margin-equivalency-for-stages} with $m\geq 2$. Then
  assuming independent priors between the situations for the
  associated finest-partition CEG $C_0$ of $C$,
  $\boldsymbol{\pi}_{v_i}\thicksim
  \Dir(\boldsymbol{\alpha}_i)$ where
  $\boldsymbol{\alpha}_i=\left(\alpha_{i1},\dots,\alpha_{im}\right)$
  for each $v_i$, $i=1,\dots,n$. Furthermore, for both $C$ and $C_0$,
  $\boldsymbol{\pi}_u\thicksim\operatorname{Dir}(\boldsymbol{\alpha}_u)$,
  where
  $\boldsymbol{\alpha}_u=\left(\sum_i\alpha_{i1},\dots,\sum_i\alpha_{im}\right)$.
\end{theorem}

\begin{proof}
  From Theorem [5] or Corollary [7], every non-trivial floret in $C_0$
  has a Dirichlet prior on its edges, which includes in this case the
  situations $v_1,\dots,v_n$. 

  Let $\gamma_{ij}=\gamma\pi_{ij}$ for $i=1,\dots,n,\: j=1,\dots,m$
  for some $\gamma\in\mathbb{R^+}$. Then it is a well-known fact that
  $\gamma_{ij}\thicksim\operatorname{Gamma}(\alpha_{ij},\beta)$ for
  all $1\leq i\leq n,1\leq j\leq m$ for some $\beta>0$, and that $\indep_{j}\gamma_{ij}$. As
  $\indep_{i}\boldsymbol{\pi}_{v_i}$, $\indep_{ij}\gamma_{ij}$. Then
  by Lemma \ref{lemma:partition-is-dirichlet}, letting $I[j]$ be the
  set of edges $\left\lbrace e_{ij}=e(v_i,v_{ij}),i=1,\dots,n\right\rbrace $ for
  $j=1,\dots,m$,
  \[
  \boldsymbol{\pi}_u\thicksim\Dir (\sum_i\alpha_{i1},\dots,\sum_i\alpha_{im})
  \]

  By margin equivalency, $\boldsymbol{\pi}_u$ must be set the same way
  for $C$.
\end{proof}

Note that the posterior of
$\boldsymbol{\pi}_u$ for a stage $u$ that is composed of the $C_0$
situations $v_1,\dots,v_n$ is thus $\boldsymbol{\pi}_u | \boldsymbol{x} \sim
\Dir (\boldsymbol{\alpha}_u^*)$ where
$\boldsymbol{\alpha}_u^*=\boldsymbol{\alpha}_u+\boldsymbol{x}_u=\sum_{i=1}^{n}{\boldsymbol{\alpha}_{v_{n}}}+\sum_{i=1}^{n}{\boldsymbol{x}_{v_{n}}}$. Equation
(\ref{eq:log-Bayes-factor}), therefore, becomes

\begin{multline}
  \log{\frac{q(C_1|\boldsymbol{x})}{q(C_2|\boldsymbol{x})}} = \log{q(C_1)}-\log{q(C_2)}+s(\boldsymbol{\alpha}_{1a})-s(\boldsymbol{\alpha}_{1a}^*)+t(\boldsymbol{\alpha}_{1a}^*)-t(\boldsymbol{\alpha}_{1a}) \\
  {}+ s(\boldsymbol{\alpha}_{1b})-s(\boldsymbol{\alpha}^*_{1b})+t(\boldsymbol{\alpha}^*_{1b})-t(\boldsymbol{\alpha}_{1b})-s(\boldsymbol{\alpha}_{1a}+\boldsymbol{\alpha}_{1b}) \\
  {}+s(\boldsymbol{\alpha}_{1a}^*+\boldsymbol{\alpha}^*_{1b})-t(\boldsymbol{\alpha}_{1a}^*+\boldsymbol{\alpha}^*_{1b})+t(\boldsymbol{\alpha}_{1a}+\boldsymbol{\alpha}_{1b})
\end{multline}

\subsection{The algorithm}
\label{sec:the-algorithm}

The algorithm thus proceeds as follows: 

\begin{enumerate}
\item Starting with the initial ET model, form the CEG $C_0$ with the finest
  possible partition, where all leaf nodes are
  placed in the terminal stage $u_\infty$ and all nodes with only one
  emanating edge are placed in the same stage. Calculate $\log q(C_{0}|\boldsymbol{x})$ using (\ref{eq:posteriorC}).
\item For each pair of situations $v_i,v_j\in C_0$ with the same number of edges, calculate $\log{\frac{q(C_1^*|\boldsymbol{x})}{q(C_0|\boldsymbol{x})}}$ where $C_1^*$ is the CEG formed by having $v_i,v_j$ in the same stage and keeping all others in their own stage; do not calculate if $q(C_1^*)=0$.
\item Let $C_1=\max_{C_{1}^*}(\log{\frac{q(C_1^*|\boldsymbol{x})}{q(C_0|\boldsymbol{x})}})$.
\item Now calculate $C_2^*$ for each pair of stages in $C_1$ except where $q(C_2^*)=0$, and record $C_2=\max(q(C_2^*|\boldsymbol{x}))$.
\item Continue for $C_3$, $C_4$ and so on until the coarsest partition $C_{\infty}$ has been reached.
\item Find $C=\max(C_0,C_1,\dots,C_{\infty})$, and select this as the MAP model.
\end{enumerate}

We note that the algorithm can also be run backwards, starting from $C_{\infty}$ and
splitting one cluster in two at each step. This has the advantage of
making the identification of positions in the MAP model easier.

\section{Examples}
\label{sec:Actual-example}

\subsection{Simulated data}
\label{sec:simulated-data}

\begin{figure}
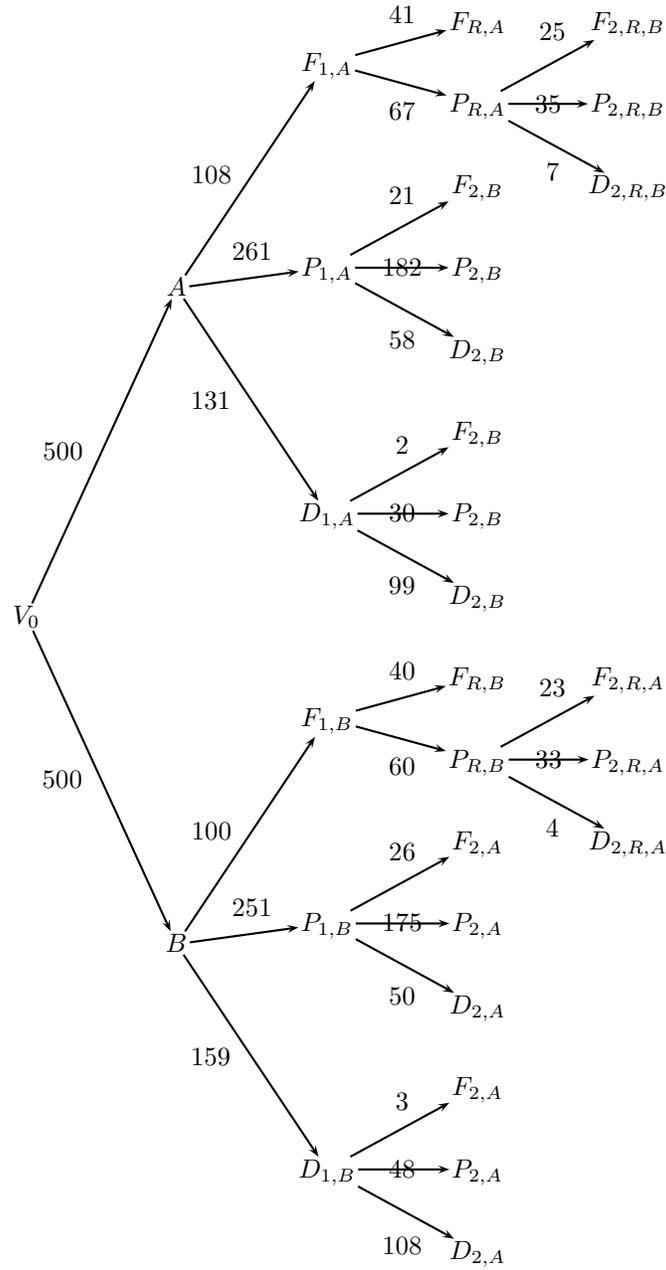

  \begin{center}

    \psset{arrows=->}
    \pstree[treemode=R,nodesep=1pt]{\TR{$V_0$}}{\pstree{\TR{$A$}\tlput{500}
      } { \pstree{
          \TR{$F_{1,A}$}\tlput{108}}{{\TR{$F_{R,A}$}\taput{41}}
          \pstree{\TR{$P_{R,A}$}\tbput{67}}{\TR{$F_{2,R,B}$}\taput{25}
            \TR{$P_{2,R,B}$}\ncput{35} \TR{$D_{2,R,B}$}\tbput{7}}}
        \pstree{\TR{$P_{1,A}$}\taput{261}}{\TR{$F_{2,B}$}\taput{21}
          \TR{$P_{2,B}$}\thput{182} \TR{$D_{2,B}$}\tbput{58}}
        \pstree{\TR{$D_{1,A}$}\tlput{131}}{\TR{$F_{2,B}$}\taput{2}
          \TR{$P_{2,B}$}\thput{30} \TR{$D_{2,B}$}\tbput{99}} }
      \pstree{\TR{$B$}\tlput{500}}{\pstree{\TR{$F_{1,B}$}\tlput{100}}{
          {\TR{$F_{R,B}$}\taput{40}} \pstree{\TR{$P_{R,B}$}\tbput{60}}
          {\TR{$F_{2,R,A}$}\taput{23} \TR{$P_{2,R,A}$}\ncput{33}
            \TR{$D_{2,R,A}$}\tbput{4}} }
        \pstree{\TR{$P_{1,B}$}\taput{251}}{\TR{$F_{2,A}$}\taput{26}
          \TR{$P_{2,A}$}\thput{175} \TR{$D_{2,A}$}\tbput{50}}
        \pstree{\TR{$D_{1,B}$}\tlput{159}}{\TR{$F_{2,A}$}\taput{3}
          \TR{$P_{2,A}$}\thput{48} \TR{$D_{2,A}$}\tbput{108}} } }

    \caption{The event tree from Example \ref{ex:education-event-tree}
      with the numbers representing the number of students in a
      simulated sample who reached each situation.}
  \end{center}
  \label{fig:education-event-tree-with-data}
\end{figure}

To first demonstrate the efficacy of the algorithm described above we
implement the algorithm using simulated data for Example
\ref{ex:education-event-tree}, where the CEG generating the data was
as known and described in Section \ref{sec:Introduction}. Figure
\ref{fig:education-event-tree-with-data} shows the number of students
in the sample who reached each situation in the tree.

In this complete dataset the progress of 1000 students has
been tracked through the event tree. Half are assigned to take module $A$
first and the other half $B$. By finding the MAP CEG model in
the light of this data we may find out whether the three hypotheses
posed in the introduction are valid. We repeat them here for
convenience:

\begin{enumerate}
\item The chances of doing well in the second component are the same
  whether the student passed first time or after a resit.
\item The components $A$ and $B$ are equally hard.
\item The distribution of marks for the second component is unaffected
  by whether students passed or got a distinction for the first
  component.
\end{enumerate}

We set a uniform prior on the CEG priors and on the root-to-leaf paths
of $C_0$, the finest partition of the tree, for illustration
purposes. The algorithm is then implemented as follows.

\begin{figure}
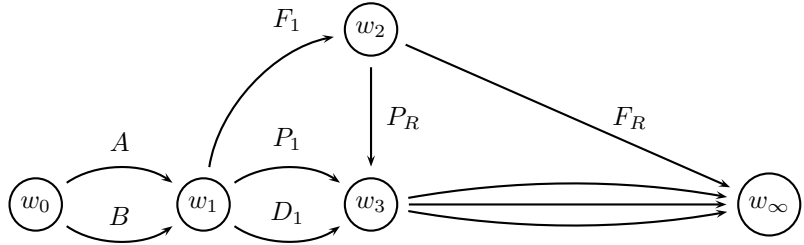
 \label{fig:education-ceg-with-data}
  \begin{center}

    $ \psmatrix[mnode=circle]
    &&w_2\\
    w_0 & w_1 & w_3 &&& w_\infty
    \endpsmatrix
    \psset{shortput=tablr,arrows=->,nodesep=4pt}
    \ncarc[arcangle=35]{2,1}{2,2}^A \ncarc[arcangle=-35]{2,1}{2,2}^B
    \ncarc[arcangle=35]{2,2}{1,3}^{F_1} \ncarc[arcangle=35]{2,2}{2,3}^{P_1}
    \ncarc[arcangle=-35]{2,2}{2,3}^{D_1} \ncline{1,3}{2,6}>{F_R}
    \ncline{1,3}{2,3}>{P_R}
    \ncarc[arcangle=10]{2,3}{2,6} \ncline{2,3}{2,6}
    \ncarc[arcangle=-10]{2,3}{2,6} $

    \caption{The MAP CEG for that event tree in Figure
    \ref{fig:education-event-tree-with-data}}
  \end{center}
\end{figure}

There are only two florets with two edges; with Beta(1,3) priors on
each and a Beta(2,6) prior on the combined stage, the log Bayes factor
is -1.85. Carrying out similar calculations for all the pairs of nodes
with three edges, it is first decided to merge the nodes $P_{1,A}$ and
$P_{1,B}$, which has a log Bayes factor of -3.76 against leaving them
apart. Applying the algorithm to the updated set of nodes and
iterating, the CEG in Figure \ref{fig:education-ceg-with-data} is
found to be the MAP one.

Under this model, it can be seen that all three hypotheses
above are satisfied and that the MAP model is the correct one.

\subsection{Student test data}
\label{sec:example2}

\begin{figure}
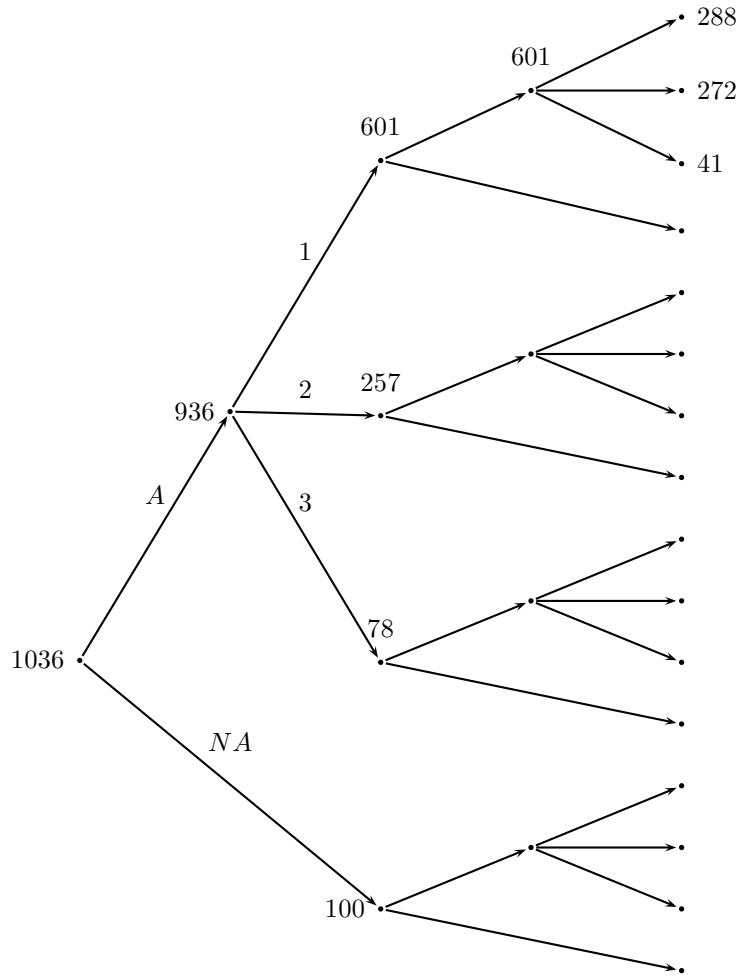

  \begin{center}

    \psset{arrows=->,radius=1pt,treemode=R,nodesep=1pt}
    \pstree{\TC*~[tnpos=l]{1036}}{ \pstree{\TC*~[tnpos=l]{936}^{$A$}}{
        \pstree{\TC*~[tnpos=a]{601}^{$1$}}{
          \pstree{\TC*~[tnpos=a]{601}}{ \TC*~{288} \TC*~{272}
            \TC*~{41} } \pstree{\Tn}{ \TC* } }
        \pstree{\TC*~[tnpos=a]{257}^{$2$}}{ \pstree{\TC*}{ \TC* \TC*
            \TC* } \pstree{\Tn}{ \TC* } }
        \pstree{\TC*~[tnpos=a]{78}^{$3$}}{ \pstree{\TC*}{ \TC* \TC*
            \TC* } \pstree{\Tn}{ \TC* } } } \pstree{\Tn}{
        \pstree{\TC*~[tnpos=l]{100}^{$NA$}}{ \pstree{\TC*}{ \TC* \TC*
            \TC* } \pstree{\Tn}{ \TC* } } } }

    \caption{Sub-tree of the event tree of possible grades for the
      MORSE degree course at the University of Warwick. Each floret of
      two edges describes whether a student's marks are available for
      a particular module (denoted by the edge labelled $A$ for the
      first module) or whether they are missing ($NA$). If they are
      available, then they are counted as grade 1 if are 70\% or
      higher, grade 2 if they are between 50\% and 69\% inclusive, and
      grade 3 if they are below 50\%. Some illustrative count data are
      shown on corresponding nodes.}
  \end{center}
  \label{fig:education-ET-2}
\end{figure}

In our second example we apply the learning algorithm to a real
dataset in order to test the algorithm's efficacy in a real-life
situation and to identify remaining issues with its usage. The dataset
we used was an appropriately disguised set of marks taken over a
10-year period from four core modules of the MORSE degree course
taught at the University of Warwick. A part of the event tree used as
the underlying model for the first two modules is shown in Figure
\ref{fig:education-ET-2}, along with a few illustrative data
points. This is a simplification of a much larger study that we are
currently investigating but large enough to illustrate the richness of
inference possible with our model search.

For simplicity, the prior distributions on the candidate models and on the
root-to-leaf paths for $C_0$ were both chosen to be uniform distributions.

The MAP CEG model was not $C_0$, so that there were some non-trivial
stages. In total, 170 situations were clustered into 32 stages.  Some
of the more interesting stages of this model are described in Table
\ref{tab:example2-map-stages}.

\begin{table}[htb]
  \centering
  \begin{tabular}{|c|c|c|c|p{1.75cm}|p{2.5cm}|}
    \hline
    Stage & Probability vector & Students & Situations & Locations & Comments \\ \hline
    7 & (0.47, 0.44, 0.08) & 685 & 2 & 1; 1,1,1 & High achievers \\ \hline
    11 & (0.22, 0.43, 0.35) & 412 & 6 & 3; 1,2; 3,1; 1,1,3 & Middling students \\ \hline
    13 & (0.33, 0.33, 0.33) & 16 & 18 & 4; 4,2; 4,3 & No students
    appeared in 17 of these situations \\ \hline
    17 & (0.07, 0.27, 0.66) & 86 & 4 & 1,3; 3,2; 3,2,4 & Struggling students \\ \hline
    27 & (0.19, 0.56, 0.25) & 464 & 7 & 1,1,4; 1,2,2; 1,3,2; 1,4,2 & More likely to get grade 2 than stage 11 \\ \hline
    28 & (0.11, 0.51, 0.38) & 436 & 6 & 1,2,3; 3,1,3; 1,2,4 & More likely to get grade 3 than stage 27 \\ \hline
  \end{tabular}
  \caption{Selected stages of MAP CEG model formed from data described in Section \ref{sec:example2}. The columns respectively detail the stage number, posterior expectation of the probability vector of that stage (rounded to two decimal places), number of students passing through that stage in the dataset, number of situations from the original ET in that stage, examples of situations in that stage (shown as sequence of grades, where ``4'' means that grade is missing), and any comments or observations related to that stage.}
  \label{tab:example2-map-stages}
\end{table}

From inspecting the membership of stages it was possible to identify
various situations which were discovered to share distributions. From
example, students who reach one of the two situations in stage 7 have
an expected probability of 0.47 in getting a high mark, an expected
probability of 0.44 of getting a middling grade, and only an expected
probability of 0.08 of achieving the lowest grade. From being in a
stage of their own, it can be deduced that students in these
situations have qualitatively different prospects from students in any
other situations. In contrast, students who reach one of the four
situations in stage 17 have an expected probability of 0.66 of getting
the lowest grade.

\section{Discussion} \label{sec:Discussion}

In this paper we have shown that chain event graphs are not just an
efficient way of storing the information contained in an event tree,
but also a natural way to represent the information that is most
easily elicited from a domain expert: the order in which events
happen, the distributions of variables conditional on the process up to
the point they are reached, and prior beliefs about the relative
homogeneity of different situations. This strength is exploited when
the MAP CEG is discovered, as this can be used in a qualitative
fashion to detect homogeneity between seemingly disparate situations. 

There are a number extensions to the theory in this paper that are
currently being pursued. These fall mostly into the two categories:
creating even richer model classes than those considered here; and
developing even more efficient algorithms for selecting the MAP model
in these model classes.

The first category includes dynamic chain event graphs. This framework
can supply a number of different model classes. The simplest case
involves selecting a CEG structure that is constant across time, but
with a time series on its parameters. A bigger class would allow the
MAP CEG structure to change over time. These larger model classes
would clearly be useful in the educational setting considered in this
paper, as they would allow for background changes in the students'
abilities, for example. 

Another important model class is that which arises from uncertainty
about the underlying event tree. A similar model search algorithm to
the one described in this paper is possible in this case after setting
a prior distribution on the candidate event trees.

In order to search any of these model classes more effectively, the
problem of finding the MAP model can be reformulated as a weighted
MAX-SAT problem, for which algorithms have been developed. This
approach was used to great effect for finding a MAP BN by Cussens
\cite{cussens_bayesian_2008}.

\section*{Appendix}
\label{sec:appendix}

Theorem \ref{theorem:dirichlet-for-c0} is based on three well-known
results concerning properties of the Dirichlet distribution, which we
review below.

\begin{lemma}
\label{lemma:dirich-is-gamma-proportions}
  Let $\gamma_j\thicksim \operatorname{Gamma}(\alpha_j,\beta), j=1,\dots,n$ where
  $\alpha_j>0$ for $j\in \lbrace 1,\dots,n\rbrace $, $\beta>0$ and
  $\indep\limits_{i\in\lbrace 1\dots n\rbrace } \gamma_i$. Furthermore, let
  $\theta_j=\frac{\gamma_j}{\gamma}$ for $j\in \lbrace 1,\dots,n\rbrace $, where
  $\gamma=\sum_{i=1}^n{\gamma_i}$.

  Then $\boldsymbol{\theta}=\operatorname*{\left(\theta_i \right)}_{i=\lbrace 1,\dots,n\rbrace } \thicksim
  \Dir \left(\alpha_1,\dots,\alpha_n\right)$.
\end{lemma}

\begin{proof}
  Kotz et al \cite{kotz_continuous_2000-2}.
\end{proof}

\begin{lemma}
\label{lemma:partition-is-dirichlet}
  Let $I[j] \subseteq \lbrace 1,\dots,n\rbrace $, $\gamma(I[j])=\sum_{i\in I[j]}\gamma_i$ and $\theta(I[j])=\sum_{i\in I[j]}\theta_i$.

  Then for any partition $I=\lbrace I[1],\dots,I[k]\rbrace $ of $\lbrace 1,\dots,n\rbrace $,
  \[
  \theta(I)=(\theta(I[1]), \theta(I[2]), \dots, \theta(I[k]))
  \thicksim
  \Dir \left(\alpha(I[1]),\dots,\alpha(I[k])\right)
  \]
  where $\alpha(I[j])=\sum_{i\in I[j]}{\alpha_i}$.
\end{lemma}

\begin{proof}
  For any $I[j] \subseteq \lbrace 1,\dots,n\rbrace $, $\indep\limits_{i \in
    I[j]}\gamma_i$, $\gamma(I[j]) \thicksim
  \operatorname{Gamma}{\left(\alpha(I[j]),\beta\right)}$ (a well-known
  result; see, for example, Weatherburn \cite{weatherburn_first_1949}), and for any
  partition $I=\lbrace I[1],\dots,I[k]\rbrace $ of $\lbrace 1,\dots,n\rbrace $, $\indep\limits_{i \in
    \lbrace 1,\dots,k\rbrace }\gamma(I[j])$. Therefore, as 
  \[
  \theta(I[j])=\sum_{i \in I[j]}{\theta_i}=\sum_{i \in I[j]}{\frac{\gamma_i}{\gamma}}=\frac{\gamma(I[j])}{\gamma}, \quad j=1,\dots,k
  \] 
  and $\gamma=\sum_{i=1}^k{\gamma(I[i])}$, the result follows from Lemma \ref{lemma:dirich-is-gamma-proportions}.
\end{proof}

\begin{lemma}
\label{lemma:floret-is-dirichlet}
  For any $I[j]\subseteq \lbrace 1,\dots,n\rbrace $ where $\left|I[j]\right|\geq 2$,
  \[
  \theta_{I[j]}=\left(\frac{\theta_i}{\theta(I[j])}\right)_{i \in
    I[j]} \thicksim \Dir \left( (\alpha_i)_{i \in I[j]} \right)
  \]
\end{lemma}

\begin{proof}
  Wilks \cite{wilks_mathematical_1962}.
\end{proof}

\begin{theorem}
\label{theorem:dirichlet-c0-appendix}
  Let the rates of units along the root-to-leaf paths $\lambda_i \in
  \Lambda, i \in \lbrace 1,\dots,\left|\Lambda \right| \rbrace $ of an event tree
  $T$ have independent Gamma distributions with the same scale
  parameter, i.e. $\gamma_i = \gamma(\lambda_i) \thicksim
  \operatorname{Gamma}(\alpha_i,\beta), i \in \lbrace 1,\dots,\left|\Lambda
  \right| \rbrace $ and $\indep\limits_{i \in
    \lbrace 1,\dots,\left|\Lambda\right|\rbrace } \gamma_i$. Then the distribution on each floret in the tree will be Dirichlet.
\end{theorem}

\begin{proof}
  Consider a floret $\mathcal{F}$ with root node $v$ and edge set
  $\lbrace e_1,\dots,e_l\rbrace $. The rate for each edge $e_i$, $\gamma(e_i)$, is
  equal to $\gamma(\lambda_{e_{i}})$, where $\lambda_{e_{i}}$ is the
  root-to-leaf path that intersects with $e_i$, so that $\gamma(e_i)
  \thicksim \operatorname{Gamma}(\alpha_{e_{i}},\beta)$ and
  $\indep\limits_{i \in \lbrace 1,\dots,l\rbrace }\gamma(e_i)$. 

  Let $I=\lbrace I[\mathcal{F}],I[\mathcal{\overline{F}}]\rbrace $ partition
  $\Lambda$, where
  $I[\mathcal{F}]=\lbrace \lambda_{e_{1}},\dots,\lambda_{e_{l}}\rbrace $ and
  $I[\mathcal{\overline{F}}]=I-I[\mathcal{F}]$. Then by Lemma
  \ref{lemma:floret-is-dirichlet}, the probability vector on
  $\mathcal{F}$ is Dirichlet, where
  \[
  \theta_{I[\mathcal{F}]}\thicksim \Dir \left( (\alpha_{e_{i}})_{i \in \lbrace 1,\dots,l\rbrace} \right)
  \]\end{proof}

\bibliographystyle{elsarticle-num}
\bibliography{library}

\end{document}